# The Contemporary Understanding of User Experience in Practice

Stefan Hellweger[1], Xiaofeng Wang[1], Pekka Abrahamsson[1]

[1]Free University of Bolzano, Bolzano, Italy
Stefan.Hellweger@stud-inf.unibz.it,{Xiaofeng.Wang,
Pekka.Abrahamsson}@unibz.it

**Abstract.** User Experience (UX) has been a buzzword in agile literature in recent years. However, often UX remains as a vague concept and it may be hard to understand the very nature of it in the context of agile software development. This paper explores the multifaceted UX literature, emphasizes the multi-dimensional nature of the concept and organizes the current state-of-the-art knowledge. As a starting point to better understand the contemporary meaning of UX assigned by practitioners, we selected four UX blogs and performed an analysis using a framework derived from the literature review. The preliminary results show that the practitioners more often focus on interaction between product and user and view UX from design perspective predominantly. While the economical perspective receives little attention in literature, it is evident in practitioners' writings. Our study opens up a promising line of request of the contemporary meaning of UX in practice.

## 1 Introduction

User experience (UX) has been a frequently discussed topic in agile literature in recent years. Especially the integration of user experience design and agile methods has caught the attention of agile researchers and practitioners alike (e.g., [1], [2], [3] and [4]). A common perception of UX in agile literature goes with the idea of separating UX designers from developers [1], [5]. The experience of the two separate teams was quite different and mainly influenced by the values of the management external to the teams [2]. More integrated approachs were proposed by adding UX designers as team members in a scrum team [3], or adding a second usability product owner in order to give UX an important role and focus [6].

However, what Don Norman, the inventor of the term *user experience*, commented about 15 years ago is still valid today: "*I invented the term because I thought human interface and usability were too narrow. I wanted to cover all aspects of the person's experience with the system including*

*industrial design, graphics, the interface, the physical interaction, and the manual. Since then the term has spread widely, so much so that it is starting to lose it's meaning... People use them often without having any idea why, what the word means, its origin, history, or what it's about."* Agile literature suffers from the same symptom. Without an explicitly and clearly defined defintion of UX, the basis for either separating UX design from or integrating it into agile development teams is not well grounded.

This observation motivated the work presented in this paper. Rather than attempting to unify different definitions of UX artificially, we admit the multi-dimensional, multi-faceted nature of the term in literature and attempt to understand how it is perceived in practice. The overall research question that drives our study is: *what is the contemporary understanding of user experience in practice?* Drawing upon the review of a set of studies that contain the definitions of UX, we are constructing a conceptual framework of UX, which acts as a sense-making device to analyze the opinions of practitioners on UX. In this paper, we present the high-level elements of the framework and the results obtained from an initial analysis of 173 blog entries selected from 4 popular UX blogs.

The rest of the paper is organized as follows: Section 2 reviews the literature from different fields that have defined UX. A synthesized view of these definitions is provided in this section. The framework was used to analyze practitioners' opinions on the same concept and the preliminary findings are reported in Section 4 after the blog analysis process is explained in Section 3. Section 5 concludes with the future work within this study.

## 2 Multiple Dimensions of User Experience in Literature

Despite the fact that there is no consensus on the definition of UX in literature, there is a common understanding that it is a complex concept and should not be equaled to usability or user interface simply. Folstad and Rolfsen [7] discover that the literature on UX may be divided in three 'camps' in terms of the relation to usability: UX encompasses usability, UX complements usability, and UX is one of several components constituting usability. For example, Hassenzahl et al. [8] argue that, instead of merely making a software usable, an expanded perspective on usability would advance the designing of user experience. Being both usable and interesting, a software system might be regarded as appealing and as a consequence the user may enjoy using it. Stage [9] argues that the recent advent of systems are focusing more on amusement and entertainment and less on work in the traditional sense, which has led some to suggest a broader notion of usability with a significantly stronger focus on UX. Based on their previous work, Hassenzahl et al. [10] summarize important distinctions between the traditional view of usability and UX. They argue that UX takes a more holistic approach, aiming for a balance between pragmatic aspects and other non-task related aspects (hedonic) of product possession and use, such as beauty, challenge, stimulation, or self-expression. In addition, UX augments the

"subjective." It is explicitly interested in the way people experience and judge products they use. What's more, UX is a more positive quality. Usability as a quality equals the removal of potential dissatisfaction. But even the best usability may never be able to "put a smile on users' faces." UX on the other hand addresses both, dissatisfiers and satisfiers, on an equal footing. The shift of emphasis from usability to experiential factors has forced researchers to consider what UX actually is and how to evaluate it [11].

Three dimensions of UX are most often suggested in the reviewed literature: user, product and interaction. As Forlizzi and Ford [12] sugget, a simple way to think about what influences experience is to think about the components of a user-product interaction, and what surrounds it. Arhippainen and Tähti [13] define UX as the experience that a person gets when he/she interacts with a product in particular conditions. The user and the product interact in the particular context of use that social and cultural factors are influencing. The user has the aspects including values, emotions, expectations and prior experience. The product has influential factors, for example, mobility and adaptivity. All these factors influence the experience that user-product interaction evokes. Similarly, Forlizzi and Battarbee [14] admit that understanding UX is complex. Designing the UX for interactive systems is even more complex, particularly when conducted by a team of multidisciplinary experts. They find that some approaches take the perspective of the user, others attempt to understand experience as it relates to the product, and a third group attempts to understand UX through the interaction between user and product. In one of the most cited UX papers, Hassenzahl and Tractinsky [15] emphasize again these three dimensions. They define UX as a consequence of a user's internal state (predispositions, expectations, needs, motivation, mood, etc.), the characteristics of the designed system (e.g. complexity, purpose, usability, functionality, etc.) and the context (or the environment) within which the interaction occurs (e.g. organizational/social setting, meaningfulness of the activity, voluntariness of use, etc.). Roto [16] takes the three components defined in [15] as a starting point and, with the knowledge on mobile browsing UX, identifies a set of attributes applicable for a wide range of UX cases.

There are also other proposals in terms of the UX dimensions, even though much less dominant. For example, in [9] UX is redefined in terms of four factors where usability is one, and the others are: branding, functionality and content. It can be argued that this redefinition reflects a more product-focused approach to UX. Oygur and McCoy [17] suggest that UX is composed of tangible (e.g., physical needs, space requirements, ergonomic issues) and intangible (e.g., emotional needs, values) aspects.

UX can be approached in a more interdisciplinary manner [14]. There are quite diverse disicplines that enable different perspectives on UX. Broadly speaking the three main perspectives are IT, design and psycology. As observed by Vliet and Mulder [18], the discussion on human experience has a long (philosophical) tradition, further explored by psychologists, neurologist and others in the last centuries up until the current time. However this vast legacy of research on human experience has for a large part not found its way into current literature on Human-Computer Interaction, Interaction Design and Usability Engineering when addressing UX. Karapanos et al. [19] discuss two threads in the UX research. One has its roots in pragmatist philosophy and the other in social psycology. More and more studies emphasize on the non-instrumental aspect of UX and delve into understanding the physio, socio, psycho and ideo needs of human beings (e.g. [20], [21]). Different perspectives and their synergy can lead to a deeper understanding of UX and their elements.

In summary, based on the literature review we built an initial conceptual framework, as shown in Table 1. The three UX elements are listed in the first column, and the three prespectives are listed in the first row. The intersection of an element and a perspective contains the sub-elements of UX from that particular perspective. The sub-elements listed in Table 1 only illustrate the conceptual framework we intend to complete. They are not exhaustive due to the preliminary phase of our research.

**Table 1: UX Elements and Perspectives**

| Perspectivs / Elements | Design | IT | Psychology |
|---|---|---|---|
| Product | novelty | mobility, adaptivity | hedonic, embodied values |
| User | ergonomic issues | quality in use, usefulness | needs, self-expression |
| Interaction | branding | rich engaging | satisfying, rewarding, emptionally fulfilling |

## 3   Research Methodology

To investigate how user experience is viewed and understood in practice, as the first step of our research we conducted a qualitative analysis of 4 most popular UX blogs written by practitioners. Blogs are considered good sources for extracting meaningful knowledge, automating trend discovery, and identifying opinion leaders [22], [23]. Therefore we believe they are valid data sources for the purpose of this study.

To sample the blogs to analyze, we first conducted a search in Goolge using the keyword "User Experience Blog". An analysis of the top search results produced a basic list of 22 candidate blogs. The list was then reviewed to remove inactive blogs that showed less then one entry in two months in average in the last two years. The blogs that did not represent the personal opinion of the author were also excluded. To ensure that the obtained sample represented the opinions of practitioners in the field we also verified that the authors of the final list of blogs have no publications in scientific journals. The final list of blogs is shown in Table 2.

The time range used to sample the entries from each blog is from January 2012 to October 2013 when the study started. In total 173 blog entries were sampled. Each entry was retrieved from the website, recorded and managed in spreadsheets. The high-level elements shown in Table 1 were used to analyze the understanding of UX reflected in these blog entries. The initial findings are reported in the next section.

**Table 2: The 4 Selected Blogs**

| Blog Name | Blog_JP | Blog_CC | Blog_MA | Blog_DA |
|---|---|---|---|---|
| URL | bokardo.com | inspireux.com | konigi.com | darmano.typepad.com/logic_emotion |
| Author | Joshua Porter | Catriona Cornett | Michael Angeles | David Armano |
| Background | Interface designer, presenter, and writer | UX designer with 6 years of experience | UX Director, former information architect | Global strategy director of a consulting company |
| Main Topics | The perception of users and how to document it. | A blog about technics for UX and technology | UX and it's relation to quality and philosophy | UX related to economy and the importance of communication |
| Starting Date | 2003 | 2008 | 2007 | 2006 |
| Entries Analysed | 73 | 15 | 36 | 49 |

## 4 User Experience from Practitioners' Perspectives

Table 3 shows the classification result of the 173 blog entries according to the framework in Table 1. As shown in the table, interaction is the most discussed dimension (87 blog entries) followed by product (65). The user dimension receives the least attention (21). The interaction dimension was especially exhaustive in these blog entries. Many aspects on the relation between product and user are discussed. The considerations cover high-level concepts such as the message the use of a product should send to its users, learnability of interaction, repositioning of elements and their different behaviors on mobile devices and desktops.

**Table 3: Classification of Blog Entries by Dimensions and Desciplines**

| Perspectivs / Elements | Design | IT | Psychology | Economy | *Total* |
|---|---|---|---|---|---|
| Product | 43 | 7 | 7 | 8 | 65 |
| User | 3 | 1 | 15 | 2 | 21 |
| Interaction | 34 | 16 | 18 | 19 | 87 |
| *Total* | 80 | 24 | 40 | 29 | 173 |

In terms of the perspectives taken, IT, design and psycology perspectives are manifested in the analyzed blog entries. One new perspective emerging from the analysis is economy. The entries classified under the "Economy" category highlight the competitive advantage on the market rising by putting weight on UX in product development.

80 blog entries are classified under the Design category. Psycology is the second biggest category (40) followed by Economy (29) and IT (24). Given the design background of the blog authors (see Table 2), it is not surprising that the main topics of the blog entries are from the design perspective. Techniques and methods of UX design are discussed, including the methods for design testing and validation. For example, one interesting concept, "Feature Deprivation", is discussed in Blog_MA: "*Designers and developers get to measure the features by removing them, and seeing how upset their users get within a controlled group*". It is a pragmatic trial in test phase by changing or deprivating features in order to discover their real value for the user.

One point worth mentioning is that a predominant context in which UX is discussed in these blog entries is the realization of websites and mobile applications. What particular interesting is the consideration related to learnability and the initial state of an application.

Figure 1 is a more detailed categorization of the blog entries per author, which allows us to gain more insights on the opinions of the author in terms of their skills, knowledge and other professional backgrounds. Some consistent patterns can be seen across the four authors. The interaction element is a dominant topic for three out of the 4 authors. Joshua Porter is an exception with more entries on the product element, which can be understood given his education background on product design. In terms of the perspectives, the blog entries of three authors have a primary focus on the design perspective except for David Armano who has more entries from the economy angle. This interest could be explained by his career change in 2011 to the senior director position on global strategy in a consulting company he was working for (described in one of his blog entries).

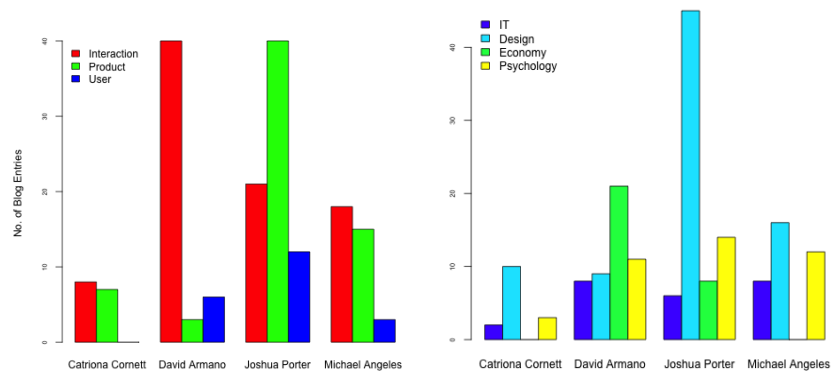

**Fig. 1: The Blog Entries Categorization Breakdown by Author**

## 5 Conclusion and Future Work

In this paper we reported the preliminary results of our research that investigates the contemporary understanding of UX in practice. In consistency with the literature, our empirical findings show that UX is also perceived as a multi-faceted phenomenon from multiple perspectives in practice. One practical implication of our study is that more integrated approaches to address UX in agile development context are in accordance with the interdisciplinary nature of UX, especialy in the context of websites and mobile application development.

In this first phase of our study we only categorized the entries using the top-level elements of the framework. The very next steps are to complete the construction of the framework with a more comprehensive list of sub-elements and then analyze the 173 blog entries using the complete framework to obtain a more comprehensive and fine-tuned picture of how user experience is understood in practice.

We will also focus on the economy perspective that emerged from the blog analysis, which does not seem to have received much attention in the relevant literature. The potential linkage to so-claimed "experience economy" [24] will be explored to better understand the economic implications of UX.